\documentstyle[aps,psfig]{revtex}

\def\e{{\mbox{e}}}
\def\esp#1{\e^{\,\textstyle#1}}
\def\bz{{\mbox{\boldmath$z$}}}
\def\bxi{{\mbox{\boldmath$\xi$}}}
\def\bE{{\mbox{\boldmath$E$}}}
\def\bF{{\mbox{\boldmath$F$}}}
\def\ba{{\mbox{\boldmath$a$}}}
\def\hba{{\hat{\ba}}}
\def\ha{{\hat{a}}}
\def\dmu#1{d\mu(#1^*\!,#1)}

\def\dmbz{\prod_\mu{ {d \bar{z}_\mu^* d \bar{z}_\mu} \over {2 \pi i} } } 
\def\H{{\cal H}}
\def\L{{\cal L}}
\def\Z{{\cal Z}}
\def\D{{\cal D}}
\def\P{{\cal P}}
\def\ave#1{\langle#1\rangle}
\def\dave#1{\langle\!\langle#1\rangle\!\rangle}

\begin{document}

%\wideabs{

\title{Effective Hamiltonian with holomorphic variables}

\author{Alessandro Cuccoli and Valerio Tognetti}

\address{Dipartimento di Fisica dell'Universit\`a di Firenze
        and Istituto Nazionale di Fisica della Materia (INFM),
        \\ Largo E. Fermi~2, I-50125 Firenze, Italy}

\author{Riccardo Giachetti}
\address{Dipartimento di Fisica dell'Universit\`a di Firenze
        and Istituto Nazionale di Fisica Nucleare (INFN),
        \\ Largo E. Fermi~2, I-50125 Firenze, Italy}

\author{Riccardo Maciocco}
\address{Dipartimento di Fisica dell'Universit\`a di Firenze,
        Istituto Nazionale di Fisica della Materia (INFM),
        \\ and Istituto Nazionale di Fisica Nucleare (INFN),
        Largo E. Fermi~2, I-50125 Firenze, Italy}

\author{Ruggero Vaia}
\address{Istituto di Elettronica Quantistica
        del Consiglio Nazionale delle Ricerche,
        via Panciatichi~56/30, I-50127 Firenze, Italy,
        \\ and Istituto Nazionale di Fisica della Materia (INFM)}

\maketitle

\begin{abstract}
The {\it pure-quantum self-consistent harmonic approximation}
(PQSCHA) permits
to study a quantum system by means of an effective classical Hamiltonian.
In this work the PQSCHA is reformulated in terms of the holomorphic variables
connected to a set of bosonic operators.
The holomorphic formulation, based on the olomorphic path integral for the
Weyl symbol of the density matrix, makes it possible to directly approach
general Hamiltonians given in terms of bosonic creation and annihilation
operators.
\end{abstract}

\vspace{1mm}

The concept of effective potential in quantum statistical mechanics
was introduced by Feynman~\cite{FH} by means of a variational
method for the  imaginary time path integral. The method was
improved by Giachetti and Tognetti~\cite{GT} and Feynman and
Kleinert~\cite{FK} using the Lagrangian path integral. Several
applications to condensed matter have demonstrated its usefulness.\cite{CGTVV}
The generalization to Hamiltonian path integrals was performed for
treating nonstandard Hamiltonians, where there is no separation
between a quadratic kinetic energy and a configurational potential,
and for which the Feynman-Jensen inequality does not hold in general.
An effective Hamiltonian for systems with non-standard Hamiltonian, as 
for ex. spin systems,
was obtained by means of the {\it pure-quantum self-consistent
harmonic approximation}, (PQSCHA).\cite{CTVV}
Nevertheless, this generalization can not be applied in a natural way
to field-theory models which are described in terms of creation and
destruction operators: the natural classical-like counterpart of this
models requires, in the Bose case, holomorphic variables.
We present here the derivation of the effective Hamiltonian in the
framework of PQSCHA for Bose systems by using an holomorphic path
integral for the Weyl symbol of the density matrix $\hat{\rho}
=\exp\{-\beta\hat{\H} \}$.

Consider a system of $N$ bosons and let $\hba = \{ \ha_\mu \}_{\mu =
1...N}$, $\hba^{\dag} = \{ \ha_\mu^{\dag} \}_{\mu = 1...N}$ be the
creation and annihilation operators. The quantum statistical average
of an operator $\hat{O}(\hba^{\dag},\hba)$ has a natural expression
in terms of the Weyl symbols of $\hat{O}$ and $\hat{\rho}$,
respectively  $O(\bz^* , \bz)$ and $\rho(\bz^* , \bz)$, 
\begin{equation}
\ave{\hat{O}} = {1\over\Z} Tr( \hat{O} \hat{\rho} )
              = {1\over\Z} \int \dmu \bz\,O(\bz^*\!,\bz)\,\rho (\bz^*\!,\bz) ~,
\label{average}
\end{equation}
where $\Z = \exp{(- \beta F)} = Tr( \hat{\rho} )$, the complex variables
$(z^*_\mu\!,z_\mu)$ clearly are the Weyl symbols of
$({\hat{a}}^{\dag}_\mu,{\hat{a}}_\mu)$, and the measure is defined as
$\dmu\bz=\prod_\mu(dz_\mu^*dz_\mu/2\pi{i})$.

The idea is to decompose the path-integral expression for
$\rho(\bz^*\!,\bz)$ into a first sum over all paths with the same average
point defined as ${1 \over \beta} \int_0^\beta { du~ \left( \bz^*(u),\bz(u) \right) }$,
and a second sum over average points. For this we introduce in the
path-integral a resolution of the identity that fixes the average
point to $(\bar{\bz}^*\!,\bar{\bz})$ and we split the integration over the
latter, defining the reduced density
\begin{equation}
 \bar{\rho} (\bz^*\!,\bz;\bar{\bz}^*\!,\bar{\bz}) =
 \int \D\big[\bz^*(u),\bz(u)\big]
 \delta \bigg( (\bar{\bz}^*\!,\bar{\bz})  - {1 \over \beta}
 \int_0^\beta {du \big(\bz^*(u),\bz(u)\big) } \bigg)
 \esp{S[\bz^*(u),\bz(u)]} ~,
\label{rhorid}
\end{equation}
where the Euclidean action $S[\bz^*(u),\bz(u)]$ is given by
\cite{BER}
\begin{eqnarray}
 S[\bz^*(u),\bz(u)] &=&
 \int_0^\beta{ du \left[ {1 \over 2} \left[ {}^t\!\dot{\bz^*}(u) \bz(u)
 - {}^t\!\bz^*(u) \dot{\bz}(u) \right] - \H(\bz^*(u),\bz(u)) \right] }
\nonumber \\
 & & - {1 \over 2} \left[ {}^t\!\bz^*(0) \bz(\beta) - {}^t\!\bz^*(\beta)
 \bz(0) \right] +
 - \left[  {}^t\!{[ \bz^*(\beta) - \bz^*(0) ]} \bz -  {}^t\!\bz^* [
 \bz(\beta)-\bz(0)]\right] ~.
 \label{action}
\end{eqnarray}

We take $\bar{\rho} (\bz^*\!,\bz;\bar{\bz}^*\!,\bar{\bz})$ as an unnormalized
probability distribution in the variables $(\bz^*\!,\bz)$ and define its
normalization constant as $\exp(- \beta \H_{\rm eff}(\bar{\bz}^*\!,\bar{\bz}))$,
so that
\begin{equation}
\bar{\rho} (\bz^*\!,\bz;\bar{\bz}^*\!,\bar{\bz}) =
 \esp{- \beta \H_{\rm eff}(\bar{\bz}^*\!,\bar{\bz})} \,
 \P(\bz^*\!,\bz;\bar{\bz}^*\!,\bar{\bz})~,
\end{equation}
and the average of $\hat{O}$ can then be written
\begin{equation}
 \ave{O} = {1 \over \Z}\! \int\!{ \dmbz \left[ \int\!{ \dmu{\bz} \,
 O (\bz^*\!,\bz)\, \P(\bz^*\!,\bz;\bar{\bz}^*\!,\bar{\bz}) } \right]
 \esp{- \beta \H_{\rm eff}(\bar{\bz}^*\!,\bar{\bz})} } \,.
\end{equation}
It is natural to interpret  $\exp(- \beta \H_{\rm eff}(\bar{\bz}^*\!,\bar{\bz}))$
as a classical-like effective density, whereas the probability
distribution $\P(\bz^*\!,\bz;\bar{\bz}^*\!,\bar{\bz})$  describes the particle
fluctuations around the point $(\bar{\bz}^*\!,\bar{\bz})$. In the classical
limit it can be seen that $\P(\bz^*\!,\bz;\bar{\bz}^*\!,\bar{\bz}) \rightarrow
\delta \left( (\bz^*\!,\bz) - (\bar{\bz}^*\!,\bar{\bz}) \right) $ and
$\exp(- \beta \H_{\rm eff}(\bar{\bz}^*\!,\bar{\bz}))$ tends to the classical Boltzmann
factor; it follows that the probability $\P$ describes the
pure-quantum fluctuations of the particle thus providing a separation
between classical-like and pure-quantum contribution to
$\ave{\hat{O}}$.

The evaluation of the reduced density $\bar{\rho}
(\bz^*\!,\bz;\bar{\bz}^*\!,\bar{\bz})$ will be done in a self-consistent
approximation
replacing $\H (\bz^*(u),\bz(u))$ in the action (\ref{action}) with a trial
Hamiltonian quadratic in the displacements from the average point,
namely
\begin{equation}
 \H_0(\bz^*\!,\bz;\bar{\bz}^*\!,\bar{\bz}) =
 {}^t\!{\bxi^*} \bE(\bar{\bz}^*\!,\bar{\bz}) \bxi +
 {1 \over 2} [ {}^t\!{\bxi} \bF (\bar{\bz}^*\!,\bar{\bz}) \bxi + c.c. ]
 + w(\bar{\bz}^*\!,\bar{\bz})\,.
\label{h0many}
\end{equation}
where $(\bxi^*\!,\bxi)=(\bz^*{-}\bar{\bz}^*\!,\bz{-}\bar{\bz})$ and
$\bE=\{E_{\mu\nu}\}$, $\bF=\{F_{\mu\nu}\}$, and $w$ are parameters
depending on $(\bar{\bz}^*\!,\bar{\bz})$ that are to be optimized.
The evaluation of $\bar{\rho}
(\bz^*\!,\bz;\bar{\bz}^*\!,\bar{\bz})$ is done by diagonalisation of the
quadratic term,
\begin{equation}
 \sum_{\mu \nu}{ \left[
 \xi_\mu^* E_{\mu \nu} \xi_\nu + {1 \over 2} ( \xi_\mu F_{\mu \nu}
 \xi_\nu + c.c. )
 \right] } = \sum_k{\omega_k
 \tilde{\xi}_k^*  \tilde{\xi}_k } ~,
\label{diagmd}
\end{equation}
with a canonical transformation $(\bxi^*\!,\bxi) \rightarrow
(\tilde{\bxi}^*\!,\tilde{\bxi})$ that, thanks to the use of path
integral with Weyl symbols, preserves at the same time the
functional measure and the form of the action.

The explicit result for $\P_0$ and $\H_{\rm eff}$ is:
\begin{equation}
 \P_0(\bz^*\!,\bz;\bar{\bz}^*\!,\bar{\bz}) = \prod_k \left( {2 \over \L(f_k)}
 \esp{ - 2 \, \tilde{\xi}_k^* \tilde{\xi}_k/\L(f_k) } \right)
 ~,~~~~~~~
 \H_{\rm eff} (\bar{\bz}^*\!,\bar{\bz}) = w(\bar{\bz}^*\!,\bar{\bz}) 
 + {1 \over \beta}
 \sum_k \ln {\sinh f_k \over f_k} ~,
\label{rhorid3}
\end{equation}
where $f_k(\bar{\bz}^*\!,\bar{\bz}) = \beta \omega_k(\bar{\bz}^*\!,\bar{\bz}) /2$
and $\L(f_k)=\coth f_k-f_k^{-1}$ is the Langevin function.
We will denote the $\P_0$-averages by double brackets $\dave{\,\cdot\,}$;
of course $\P_0$ turns out to be a Gaussian centered in
$(\bar{\bz}^*\!,\bar{\bz})$, so it is fully determined through its
second moments $\dave{\tilde\xi^*_k\tilde\xi_{k'}}=\delta_{kk'}\,\L(f_k)/2$,
$\dave{\tilde\xi^*_k\tilde\xi^*_{k'}}=\dave{\tilde\xi_k\tilde\xi_{k'}}=0$\,.
The linear canonical transformation made can be used to express in a simple
way the moments for the original variables, $(\xi^*_\mu,\xi_\mu)$.

In order to determine the parameters of $\H_0$ in a
self-consistent way we require that the original and the trial
Hamiltonian, and their second derivatives, have the same $\P_0$-averages:
\begin{equation}
 \dave{ \H (\bar{\bz}^*{+}\xi^*\!,\bar{\bz}{+}\xi) } =
 \dave{ \H_0 (\bar{\bz}^*{+}\xi^*\!,\bar{\bz}{+}\xi;\bar{\bz}^*\!,\bar{\bz}) }
 = \sum_k{\omega_k(\bar{\bz}^*\!,\bar{\bz})
 { {\L \left(f_k(\bar{\bz}^*\!,\bar{\bz})\right)} \over 2} }
 + w(\bar{\bz}^*\!,\bar{\bz}) \,,
\label{schmd}
\end{equation}
\begin{eqnarray}
 \dave{ \partial_{z_{\mu}^*} \partial_{z_{\nu}} \H
 (\bar{\bz}^*{+}\xi^*\!,\bar{\bz}{+}\xi) }
 &=&
 \dave{  \partial_{z_{\mu}^*} \partial_{z_{\nu}}
 \H_0 (\bar{\bz}^*{+}\xi^*\!,\bar{\bz}{+}\xi;\bar{\bz}^*\!,\bar{\bz}) }
 \equiv E_{\mu \nu}(\bar{\bz}^*\!,\bar{\bz}) \,,
\nonumber \\
 \dave{  \partial_{z_{\mu}} \partial_{z_{\nu}} \H
 (\bar{\bz}^*{+}\xi^*\!,\bar{\bz}{+}\xi) }
 &=&
 \dave{  \partial_{z_{\mu}} \partial_{z_{\nu}} \H_0
 (\bar{\bz}^*{+}\xi^*\!,\bar{\bz}{+}\xi;\bar{\bz}^*\!,\bar{\bz}) }
 \equiv F_{\mu \nu}(\bar{\bz}^*\!,\bar{\bz}) \,,
\nonumber \\
 \dave{  \partial_{z_{\mu}^*} \partial_{z_{\nu}^*} \H
 (\bar{\bz}^*{+}\xi^*\!,\bar{\bz}{+}\xi) }
 &=&
 \dave{  \partial_{z_{\mu}^*} \partial_{z_{\nu}^*} \H_0
 (\bar{\bz}^*{+}\xi^*\!,\bar{\bz}{+}\xi;\bar{\bz}^*\!,\bar{\bz}) }
 \equiv F_{\mu \nu}^*(\bar{\bz}^*\!,\bar{\bz}) \,.
\label{sce}
\end{eqnarray}
where $\dave{\cdot}$ stands for the Gaussian average
given by $\P_0$.

Finally we observe that a solution of the general problem for arbitrary
values of $(\bar{\bz}^*\!,\bar{\bz})$ is rather difficult to determine
for many degrees of freedom: a further simplification is in order.
This is the {\it low-coupling approximation}\,\cite{CGTVV} (LCA) and its
main purpose is to make the averages $\dave{\,\cdot\,}$ independent
of the phase-space point $(\bar{\bz}^*\!,\bar{\bz})$, so that the
above self-consistent equations are to be solved only once.
The simplest way, consists in expanding the matrices
$\bE(\bar{\bz}^*\!,\bar{\bz})$, $\bF(\bar{\bz}^*\!,\bar{\bz})$
around a self-consistent minimum of $\H_{\rm eff}$.
The effective Hamiltonian and the expressions found for thermal
averages are parallel to those shown in Ref.\,\cite{CGTVV}.

In perspective, one can think of translating this formalism to the fermionic
case, getting ``classical'' expressions in terms of Grassman variables.

\end{document}